\documentclass{article}
\usepackage{spconf,amsmath,graphicx,booktabs,xspace,amsfonts,cite,multirow,makecell}
\usepackage[colorlinks,linkcolor=black,anchorcolor=black,citecolor=black,urlcolor=black]{hyperref}

 \title{Zero-shot Personalized Lip-to-Speech Synthesis With face image based voice control}
 \name{Zheng-Yan Sheng, Yang Ai, Zhen-Hua Ling* \thanks{* Corresponding author. This work was partially supported by the Fundamental Research Funds for the Central Universities.}}
 \address{National Engineering Research Center of Speech and Language Information Processing
 \\University of Science and Technology of China, Hefei, P.R.China\\
 {\small \tt \ zysheng@mail.ustc.edu.cn, yangai@ustc.edu.cn, zhling@ustc.edu.cn}}

\begin{document}
\ninept
\maketitle
\begin{abstract}
Lip-to-Speech (Lip2Speech) synthesis, which predicts corresponding speech from talking face images, has witnessed significant progress with various models and training strategies in a series of independent studies. However, existing studies can not achieve voice control under zero-shot condition, because extra speaker embeddings need to be extracted from natural reference speech and are unavailable when only the silent video of an unseen speaker is given. 
In this paper, we propose a zero-shot personalized Lip2Speech synthesis method, in which face images control speaker identities.
A variational autoencoder is 
adopted to disentangle the speaker identity and linguistic content representations, which enables speaker embeddings to control the voice characteristics of synthetic speech for unseen speakers.
Furthermore, we propose associated cross-modal representation learning to promote the ability of face-based speaker embeddings (FSE) on voice control.  
Extensive experiments verify the effectiveness of the proposed method whose synthetic utterances are more natural and matching with the personality of input video than the compared methods. To our best knowledge, this paper makes the first attempt on zero-shot personalized Lip2Speech synthesis with a face image rather than reference audio to control voice characteristics.

\end{abstract}

%

\section{Introduction}

Lip-to-Speech (Lip2Speech) synthesis is a task to predict the corresponding speech given a sequence of talking face images. This task has various applications, such as assisting patients who are unable to produce voiced sounds (aphonia), restoring speech for videoconferencing in a noisy environment and dubbing silent movies. 

With the development of deep learning, mainstream Lip2Speech studies follow the encoder-decoder framework, where the encoder extracts the linguistic content and voice characteristics information from the talking video, and the decoder converts the above information to a corresponding audio. Recently, a few Lip2Speech methods \cite{he2022flow, prajwal2020learning} have been proposed for  speaker-dependent and large-vocabulary scenarios, which focused on improving the linguistic content representation and speech quality for a single seen speaker. Some other studies \cite{kim2021lip, mira2021end, vougioukas2019video, de2022svts, wang2022vcvts} attempted to achieve multi-speaker Lip2Speech for both seen and unseen speakers. Most of them utilized the speech-based speaker embeddings (SSE) extracted from natural reference speech using a speaker verification model, and then combined the output of a video encoder with SSE as the input of the decoder. In other words,  these methods still need a small amount of unseen speakers' reference speech, failing to achieve zero-shot Lip2Speech synthesis under the condition that only the silent video of a target speaker is available at the synthesis stage.

There are two main challenges to achieve zero-shot Lip2Speech synthesis with face image based voice control, i.e., 
disentangling the representations for speaker identities and linguistic contents from input videos, and enabling face images to control voice characteristics.
First, silent video input contains both linguistic content and speaker identity information. 
Several existing multi-speaker Lip2Speech methods \cite{prajwal2020learning,vougioukas2019video,kim2021lip,mira2021end} synthesized speech with brilliant voice characteristics for seen speakers, while for unseen speakers the quailty of synthetic speech  degraded significantly and the voice personality sometimes conflicted with the face personality of the input video.
Hence, it is necessary to exclude speaker identity information from the output of  the video content encoder so that the voice characteristics of synthetic speech can be fully controlled by extra speaker embeddings.
Second, due to the limited number of speakers in popular Lip2Speech datasets, existing multi-speaker Lip2Speech methods \cite{prajwal2020learning,vougioukas2019video,kim2021lip,mira2021end} usually failed to establish a stable and generalized mapping from  input face images to  voice characteristics for unseen speakers. 
Therefore, it is necessary to extract robust face-based speaker embeddings (FSE) to describe the correspondence between these two spaces.

\begin{figure*}[t]
  \vspace{-1.5em}
  \centering
  \setlength{\abovecaptionskip}{1.cm}
  \includegraphics[width=12cm]{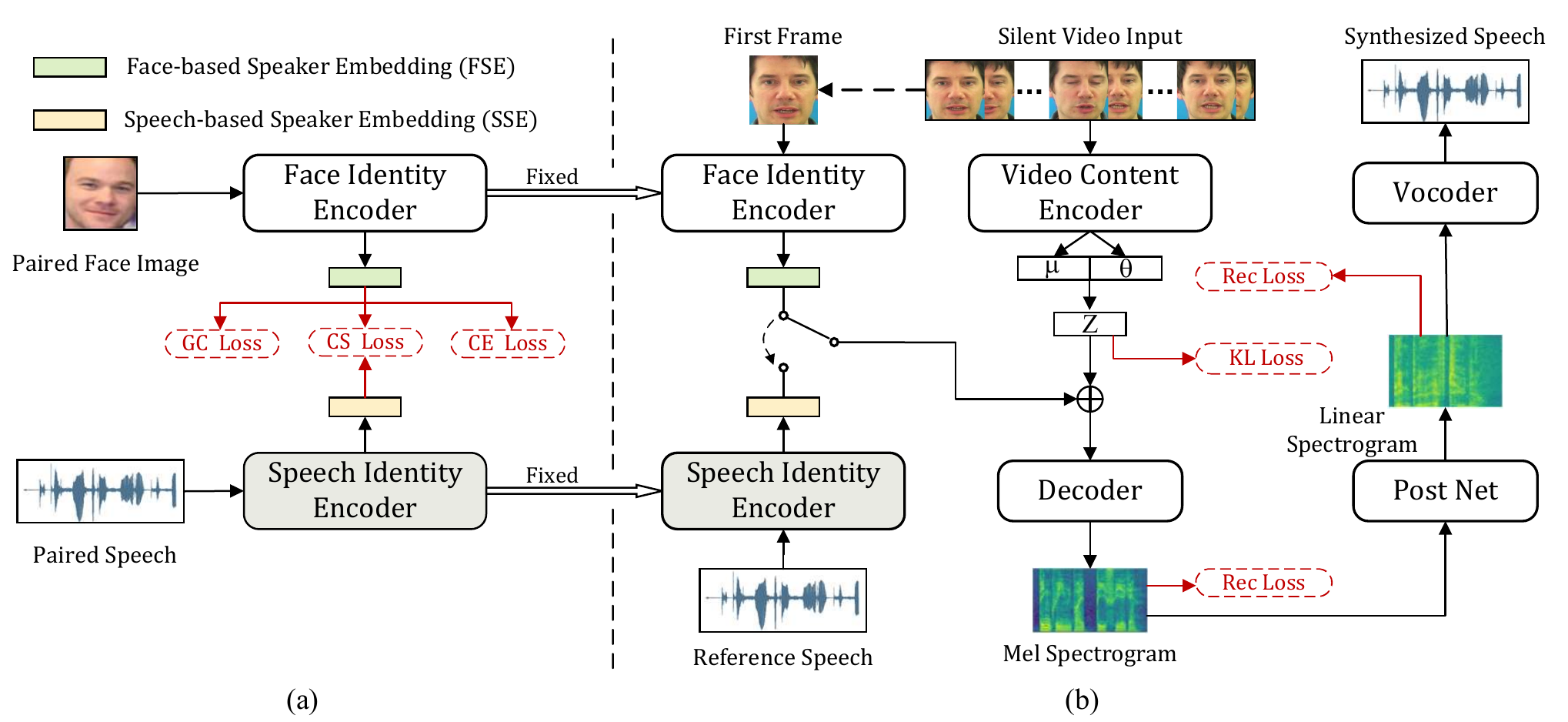}
  \label{overall}
  \caption{The overall flowchart of our proposed method containing (a) Associated Voice-Face Representation Learning and (b) Lip-to-Speech Synthesis. GC Loss, CS Loss, CE Loss represents gender contrastive loss, cosine similarity loss and cross-entropy loss, respectively. Rec Loss represents the L1 reconstruction loss for mel spectrograms and linear spectrograms, and KL loss stands for the Kullback-Leibler divergence between the posterior and prior distributions of ${z}$. 
  	Paired face image and speech are selected from the same person. The switch in (b) selects either the face identity encoder or the speech identity encoder to produce speaker embeddings for voice control at the inference stage. At the training stage, the switch always selects the face identity encoder. 
  }
   \vspace{-1.5em}
\end{figure*}

To overcome the two challenges mentioned above, we propose a zero-shot personalized Lip2Speech synthesis method, where the speaker identity and voice characteristics of synthetic speech can be controlled by an unseen speaker's face image. A variational autoencoder (VAE) is utilized as the backbone of our video content encoder to discard speaker identity  information from the input video and reserve only linguistic contents. Then the disentangled content representations and the speaker embedding given by a face identity encoder are concatenated as the decoder's input to predict the acoustic features for waveform reconstruction.  In addition, an associated voice-face representation learning strategy is proposed to train the face identity encoder.  This strategy transfers the knowledge from a speech identity encoder to the face identity encoder, enabling face images to provide necessary voice characteristics in the Lip2Speech process. Evaluation and analysis experiments on the GRID\cite{cooke2006audio} dataset demonstrate the effectiveness of our proposed method, which synthesizes more natural and personality-matching speech than other compared methods.

\vspace{-1mm}
\section{PROPOSED METHOD}
\vspace{-1mm}
The overall flowchart of our proposed method is shown in Fig. \ref{overall}, which mainly includes a face identity encoder, a speech identity encoder, a video content encoder, a decoder, a post net and a vocoder. 
At the training stage, the speech identity encoder is built with a speaker recognition task and the face identity encoder is trained by associated cross-modal representation learning.
Then, the other model components, including the video content encoder, the decoder and the post net, are trained with the fixed face identity encoder under a VAE-based  Lip2Speech synthesis framework.
At the inference stage, given the silent video input from an unseen speaker, a speaker embedding from either the speech identity encoder or the face identity encoder is adopted to control the voice characteristics of synthetic speech.
When the speech identity encoder is used, a reference utterance from the traget speaker is needed.
When the  face identity encoder is used, zero-shot personalized Lip2Speech synthesis can be achieved since no reference speech of unseen speakers is necessary any more.

\vspace{-1mm}
\subsection{VAE-based Lip2Speech Synthesis}
\vspace{-1mm}
As shown in Fig.~\ref{overall}(b), inspired by VAE-based voice conversion \cite{hsu2016voice}, we introduce the VAE framework to extract speaker-independent linguistic content representations from video input. Specifically, given a talking head video ${\boldsymbol{X}=[\boldsymbol{x}_1, \boldsymbol{x}_2, \ldots, \boldsymbol{x}_T]}$, where $T$ is the number of video frames and ${\boldsymbol{x}_t}$ represent the ${t}$-th frame in this video, 
a sequence of latent vectors ${\boldsymbol{Z}=[\boldsymbol{z}_1, \boldsymbol{z}_2, \ldots, \boldsymbol{z}_T]}$  are extracted.
The video content encoder predicts the mean ${\boldsymbol{\mu}}_t$ and the variance ${\boldsymbol{\sigma}_t^{2}}$ of the posterior distribution of $\boldsymbol{z}_t$, i.e.,  ${q_{\phi}(\boldsymbol{z}_t \mid \boldsymbol{X})=\mathcal{N}\left({\boldsymbol{\mu}_t}, \boldsymbol{\sigma}_t^{2} \mathbf{I}\right)}$. The posterior distribution is constrained by the  Kullback-Leibler (KL) divergence with the prior distribution of $\boldsymbol{z}_t$, i.e., $p(\boldsymbol{z}_t)=\mathcal{N}(\mathbf{0}, \mathbf{I})$, where 
$\mathbf{I}$ is an identity matrix meaning that all dimensions are independent of each other.  Then, the hidden representations of linguistic contents, i.e., the values of $\boldsymbol{z}_t$, can be sampled from ${q_{\phi}(\boldsymbol{z}_t \mid \boldsymbol{X})}$ by the reparameterization trick. Due to the intrinsic synchronization between the video stream and the speech stream, the input video and the output mel spectrograms are fixedly aligned. Hence, we simply upsample the hidden representations by ${\alpha}$ times to match the sampling rate of mel spectrograms, where ${\alpha}$ is the ratio between the sampling rates of mel spectrograms and video frames. Following the architecture widely used in lip reading tasks \cite{ma2022training,ma2021towards,martinez2020lipreading}, the video encoder is composed of 3D convolutional blocks, a standard 2D ResNet-18 and two convolutional layers.

At the training stage, the first frame of the video input 
is sent into the face identity encoder to extract the face-based speaker embedding (FSE) of the target speaker. 
Then, the speaker embedding and upsampled hidden representations are concatenated and further used as the input for the decoder, which adopts a non-autoregressive Conformer architecture \cite{gulati2020conformer}. 
The Conformer decoder converts the speaker embedding and linguistic content features to mel spectrograms. To further improve the quality of synthetic speech, we adopt a post net to convert the mel spectrograms  into linear spectrograms. In the end, the Griffin-Lim algorithm is used to recover phases from the linear spectrograms and to reconstruct the final waveforms.

The model is trained by optimizing the evidence lower bound (ELBO) loss, i.e., 
\begin{equation}
\mathcal{L}=\mathbb{E}_{q_{\phi}(\boldsymbol{Z}|\boldsymbol{X})}\left[\log p_{{\theta}}(\boldsymbol{Y} \mid \boldsymbol{Z}, \boldsymbol{e})\right] -\lambda D_{K L}\left(q_{\phi}(\boldsymbol{Z} \mid \boldsymbol{X}) \| p(\boldsymbol{Z})\right),
\label{eq1}
\end{equation}
where ${\boldsymbol{Y}=[\boldsymbol{y}_1, \boldsymbol{y}_2, \ldots, \boldsymbol{y}_{\alpha T}]}$ denote ground truth spectrograms and $\boldsymbol{e}$ stands for the speaker embedding. ${\phi}$ and ${\theta}$ represent the parameters of the video content encoder and other modules except the vocoder, respectively.
The first term on the left side of Eq. (\ref{eq1}) corresponds to the L1 reconstruction loss for both mel spectrograms and linear spectrograms. ${D_{K L}}$ calculates the KL divergence between the posterior and prior distributions of $\boldsymbol{Z}$. 

\vspace{-1mm}
\subsection{Associated Voice-Face Representation Learning}
\label{subsec:ARL}
\vspace{-1mm}
It is straightforward to use a pretrained face recognition model as the face identity encoder to extract FSEs. However, face recognition models lack the ability of linking face identities with voice characteristics. On the other hand, jointly training the face identity encoder with other modules in our VAE-based Lip2Speech model may not perform well for unseen speakers due to the limited number of speakers in most Lip2Speech training sets.
Therefore, we present a method of associated voice-face representation learning that enables one face image to provide its corresponding speaker's voice characteristics, as shown in Fig.~\ref{overall}(a). This method mainly contains a speech identity encoder and a face identity encoder, which are 
pretrained on a speaker verification task and a face recognition task, respectively. The speech identity encoder follows the structure proposed in utterance-level speaker recognition \cite{xie2019utterance}, which consists of a thinResNet trunk architecture and a dictionary-based NetVLAD layer to aggregate features across time. Moreover, the InceptionResNet architecture \cite{szegedy2017inception} is adopted for the face identity encoder. Then, an utterance and a face image from the same speaker are adopted as the paired input  in Fig.~\ref{overall}(a) 
for cross-modal representation learning. During the learning process, the parameters of the speech identity encoder are fixed, and the  face identity encoder is updated by a cosine similarity loss ${\mathcal{L}}_{CS}$, a gender contrastive loss ${\mathcal{L}}_{GC}$ and a cross-entropy loss ${\mathcal{L}}_{CE}$.

The cosine similarity loss ${\mathcal{L}}_{CS}$ is calculated between the SSE and FSE vectors extracted by the two identitiy encoders from the paired speech and image input. 
Considering that gender is an important voice characteristic, the gender contrastive loss \cite{hadsell2006dimensionality} ${\mathcal{L}}_{GC}$ is introduced as
\begin{equation}
\begin{aligned}
\mathcal{L}_{GC}=-\log \frac{ \sum_{{i=1}}^{N} \sum_{g(\boldsymbol{v}_{i})=g(\boldsymbol{u}_{i})} \exp \left(\boldsymbol{v}_{i}^\top  \boldsymbol{u}_{i} \right)}{ \sum_{{i=1}}^{N} \sum_{g(\boldsymbol{v}_{i}) \neq g(\boldsymbol{u}_{i})} \exp \left(\boldsymbol{v}_{i}^\top  \boldsymbol{u}_{i} \right)},\\
\end{aligned}
\label{eq2}
\end{equation}
where ${N}$ is the batch size, ${\boldsymbol{V}=[\boldsymbol{v}_1, \boldsymbol{v}_2, \ldots, \boldsymbol{v}_{N}]}$ stands for the FSE vectors in a batch and is  randomly shuffled to get ${\boldsymbol{U}=[\boldsymbol{u}_1, \boldsymbol{u}_2, \ldots, \boldsymbol{u}_{N}]}$, ${i}$ stands for the index in a batch, and $g()$ is a function to get the gender of ${\boldsymbol{v_i}}$ and ${\boldsymbol{u_i}}$. The value of the gender contrastive loss is low when the predicted FSE is similar to that of the same gender and dissimilar to that of the opposite gender. The cross-entropy loss ${\mathcal{L}}_{CE}$ for the face recognition task is also added, which keeps the discrimination ability of the face identity encoder and provides more supervision information at the training stage.

\vspace{-1mm}
\section{Experiments}
\vspace{-1mm}

\vspace{-1mm}
\subsection{Datasets}
\vspace{-1mm}
Our experiments were conducted on the GRID\cite{cooke2006audio} dataset, which consisted of 33 speakers with 1K videos per speaker. The dataset was recorded in an artifically constrained environment, and each utterance in the video contains six context independent words. 
We selected the data of 10 female speakers (s4, s7, s15, s16, s18, s22, s23, s29, s31, s34) and 10 male speakers (s3, s5, s6, s10, s12, s14, s17, s26, s28, s32) to construct the training set, the development set and the test for seen speakers. The split ratios between these three subsets were 90\%, 5\% and 5\% for each speaker. The data of the remaining 13 speakers was used as the test set for unseen speakers.

The Voxceleb2 \cite{chung2018voxceleb2} and VGGFace2\cite{cao2018vggface2} datasets were used for training the speech identity encoder and pre-training the face identity encoder. The Voxceleb2 dataset contains over 1 million utterances from 6,112 celebrities extracted from the videos uploaded to YouTube.  The VGGFace2 dataset is comprised of around 3.31 million images divided into 9131 classes, each representing a specific person. 
The speech identity encoder and the face identity encoder were firstly trained on the two datasets respectively. Then, the data of the 5,994 speakers that existed in both datasets was used for the cross-modal training. A face image and an utterance from the same speaker were randomly selected from the two datasets 
to form the paired input introduced in  Section \ref{subsec:ARL}. 

\vspace{-1mm}
\subsection{Implementation Details}
\vspace{-1mm}
To get the input of the video content encoder, each frame of videos in the GRID dataset was cropped to the center of the lip and resized to 112 ${\times}$ 112.
The frame rate of videos in the GRID dataset was 25 fps. 
To get the ground-truth ouput of the decoder, the audio in the GRID dataset was resampled to 16kHz, and converted into 80-dimensional mel spectrograms and 321-dimensional linear spectrograms by using an FFT size of 640, a Hamming window length of 640 and a hop length of 160. Thus, the frame rate ratio 
${\alpha = 4}$ in Section 2.1.
The number of Conformer layers in the decoder was 5, and each layer consisted of a 256-dimensional attention, 4 attention heads and a 31 ${\times}$ 1 convolutional kernel. The post net consisted of 5 convolution layers, which mainly contained 1D convolution, bath normalization and Leaky rectified linear unit (LReLU). The VAE-based Lip2Speech synthesis model was trained by an Adam optimizer \cite{kingma2014adam} with a learning rate of 5e-4. The weight ${\lambda}$  of the KL term was set to 0.001.

The speech identity encoder accepted 
80-dimensional mel spectrograms extracted from the audios sampled at 16kHz 
with an FFT size of 1024, a Hamming window length of 400 and a hop length of 160, following the work of utterance-level speaker recognition \cite{xie2019utterance}.
The Dlib toolkit \cite{king2009dlib} was used to obtain face landmarks for the images in VGGFace2 and the first frame of videos in GRID, and then the face images were cropped and resized to 160 ${\times}$ 160 as the input of the face identity encoder.
The SSE and  FSE  vectors were defined as the output of the penultimate linear layer of the speech identity encoder and the face identity encoder, respectively. The dimension of both embeddings were 512.
The  speech identity encoder trained on Voxceleb2\footnote{https://github.com/WeidiXie/VGG-Speaker-Recognition} was fixed,
and the face identity encoder pretrained on VGGFace2\footnote{https://github.com/timesler/facenet-pytorch} was further optimized using three loss functions by an Adam optimizer \cite{kingma2014adam} with a learning rate of 1e-3 and batch size of 512. 
The extracted 512-dimensional SSE or FSE vectors were then mapped into 256-dimensional ones through a linear layer and a rectified linear unit (ReLU) before concatenating them with the 512-dimensional linguistic content representations for decoding.

\vspace{-1mm}
\begin{table*}[]
\vspace{-1.5em}
\label{table1}
\begin{center}
\caption{Objective and subjective evaluation results of compared methods.}
\label{table1}
\setlength{\tabcolsep}{1.5mm}{
  \begin{tabular}{c|lllcc|lllccc}
    \Xhline{2\arrayrulewidth}
    \multirow{2}{*}{Method}  & \multicolumn{5}{c|}{Seen}                                                                                                  & \multicolumn{6}{c}{Unseen}                                                                                                                                \\ \cline{2-12} 
                             & \multicolumn{1}{l|}{STOI}  & \multicolumn{1}{l|}{ESTOI} & \multicolumn{1}{l|}{PESQ}  & \multicolumn{1}{c|}{EER}   & MOS-SN & \multicolumn{1}{l|}{STOI}  & \multicolumn{1}{l|}{ESTOI} & \multicolumn{1}{l|}{PESQ}  & \multicolumn{1}{c|}{EER}   & \multicolumn{1}{c|}{MOS-SN} & MOS-FVM \\ \hline
    Ground Truth (Grinfin-Lim) & \multicolumn{1}{l|}{0.802} & \multicolumn{1}{l|}{0.696} & \multicolumn{1}{l|}{3.293} & \multicolumn{1}{c|}{6.786} & 3.820  & \multicolumn{1}{l|}{0.801} & \multicolumn{1}{l|}{0.698} & \multicolumn{1}{l|}{3.289} & \multicolumn{1}{c|}{6.751} & \multicolumn{1}{c|}{3.750}  & 3.783   \\
    Proposed\_Speech         & \multicolumn{1}{l|}{0.741} & \multicolumn{1}{l|}{0.619} & \multicolumn{1}{l|}{1.914} & \multicolumn{1}{c|}{6.944} & 3.057  & \multicolumn{1}{l|}{0.598} & \multicolumn{1}{l|}{0.407} & \multicolumn{1}{l|}{1.477} & \multicolumn{1}{c|}{26.57} & \multicolumn{1}{c|}{2.830}  & 3.172   \\ \hline
    Proposed                 & \multicolumn{1}{l|}{0.739} & \multicolumn{1}{l|}{0.615} & \multicolumn{1}{l|}{1.885} & \multicolumn{1}{c|}{7.095} & 3.049  & \multicolumn{1}{l|}{0.589} & \multicolumn{1}{l|}{0.389} & \multicolumn{1}{l|}{1.454} & \multicolumn{1}{c|}{36.72} & \multicolumn{1}{c|}{2.817}  & 3.133  \\
    Proposed-VAE             & \multicolumn{1}{l|}{0.738} & \multicolumn{1}{l|}{0.619} & \multicolumn{1}{l|}{1.926} & \multicolumn{1}{l|}{6.923} & 3.246   & \multicolumn{1}{l|}{0.547} & \multicolumn{1}{l|}{0.328} & \multicolumn{1}{l|}{1.404} & \multicolumn{1}{l|}{43.13} & \multicolumn{1}{c|}{2.690}  & 2.378   \\
    Proposed-CML             & \multicolumn{1}{l|}{0.735} & \multicolumn{1}{l|}{0.610} & \multicolumn{1}{l|}{1.883} & \multicolumn{1}{c|}{7.287} & 3.012  & \multicolumn{1}{l|}{0.570} & \multicolumn{1}{l|}{0.356} & \multicolumn{1}{l|}{1.435} & \multicolumn{1}{c|}{44.71} & \multicolumn{1}{c|}{2.783}  & 2.611   \\ \Xhline{2\arrayrulewidth}
    \end{tabular}}
\end{center}
\vspace{-1.5em}
\end{table*}
\vspace{-1mm}

\vspace{-1mm}
\subsection{Comparison Systems}
\vspace{-1mm}
Five systems were compared to evaluate the performance of our proposed method, i.e., 
(1) \textbf{Ground Truth (Grinfin-Lim)}, which transferred the natural linear spectra of test utterances to waveforms using the Griffin-Lim algorithm; 
(2) \textbf{Proposed\_Speech}, which was our proposed method but utilized reference utterances of target speakers to extract speaker embeddings at the inference stage; 
(3) \textbf{Proposed},  which was our proposed method and extracted speaker embeddings from the face images  of target speakers to achieve zero-shot personalized lip-to-speech synthesis;
(4) \textbf{Proposed-VAE}, which was the same as \textbf{Proposed} except that the VAE structure was removed; 
(5) \textbf{Proposed-CML}, which was the same as \textbf{Proposed} except that the associated voice-face representation learning was not applied to train the face identity encoder.

\vspace{-1mm}
\subsection{Evaluation Results}
\vspace{-1mm}
Following previous studies \cite{kim2021lip,de2022svts}, the short-time objective intelligibility (STOI), extended short-time objective intelligibility (ESTOI) and perceptual evaluation of speech quality (PESQ) \cite{rec2005p} were employed as objective evaluation metrics. STOI and ESTOI were metrics for  speech intelligibility, and  PESQ measures the perceptual quality of speech.
We also introduced another objective metrics to evaluate the speaker similarity of synthetic speech, i.e., equal error rate (EER) \cite{chen2021towards}, which was widely used in speaker verification tasks \cite{xie2019utterance}. Lower EER indicates better similarity between the voice characteristics of the synthesized speech and that of the ground truth. For subjective evaluation, mean opinion score for speech naturalness (MOS-SN) was used to quantitatively measure the naturalness of synthetic speech, and mean opinion score for face-voice matching degree (MOS-FVM) was used to evaluate whether the face in the input video and the synthetic voice are compatible with each other.
These two subjective metrics were evaluated on Amazon Mechanical Turk platform\footnote{https://www.mturk.com/}. 
 20 sentences were randomly selected from the test set and were synthesized by the five systems.  A total of 20 listeners participated in the test, and each listener needed to give a score  from 1 (completely unnatural or completely mismatched) to 5 (completely natural or completely matched) for each synthetic utterance on each metric.  
All evaluation results on both seen and unseen test sets are reported in Table \ref{table1}.\footnote{The synthesized samples can be found at \url{https://levent9.github.io/Lip2Speech-demo/}.}

Since the face identity encoder was trained under the guidance of the speech identity encoder using the ${\mathcal{L}}_{CS}$ loss, the \textbf{Proposed\_Speech} system can be regarded as the upper bound of the \textbf{Proposed} system. Table \ref{table1} shows that \textbf{Proposed\_Speech} achieved a much lower EER than \textbf{Proposed} for unseen speakers, while the performance of these two systems was close on other metrics, which demonstrated the feasibility of face image based voice control.

We can see that the \textbf{Proposed} system outperformed the \textbf{Proposed-VAE} system on all metrics significantly (p\textless 0.05 in paired t-tests) for unseen speakers, 
especially on MOS-FVM.
A close examination found that  \textbf{Proposed-VAE} sometimes produced speech with a gender opposite to that of the input video, while \textbf{Proposed} didn't make such mistakes.
On the other hand,  \textbf{Proposed}  performed worse than  \textbf{Proposed-VAE} with a lower score in MOS-SN  and other three objective metrics (p\textless 0.05) for seen speakers, indicating a trade-off between disentanglement and the speech quality of seen speakers.
Compared with the \textbf{Proposed-CML} system, \textbf{Proposed} achieved much better performance on MOS-FVM and EER (p\textless 0.05) for unseen speakers. This indicates the effectiveness of the three loss functions that guided the face identity encoder to  produce more robust and voice-related FSE vectors.

\begin{figure}[t]
	\centering
	\setlength{\abovecaptionskip}{1.cm}
	\includegraphics[width=8cm]{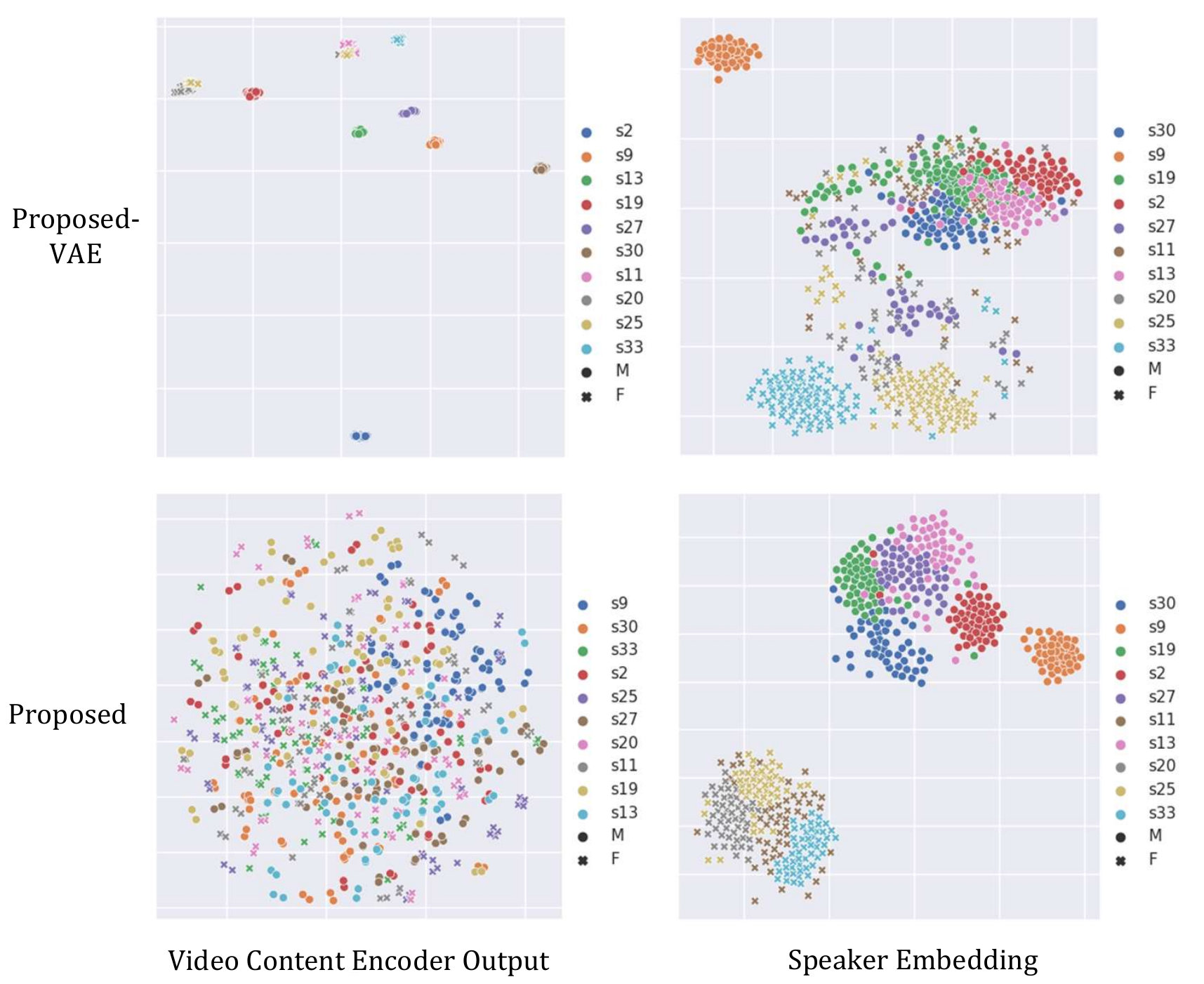}
	
	\caption{t-SNE visualization of the outputs of the video content encoder and the speaker embeddings extracted from the synthetic speech of unseen speakers. 
	Each point stands for one utterance. Colors and shapes indicate speaker identities and genders respectively.}
	\label{visual}
	\vspace{-1em}
\end{figure}

\begin{table}[t]%
  \vspace{-1.0em}
	\begin{center}
		\caption{Performance comparison on Lip2Speech synthesis with reference speech based voice control.}
		\label{table2}
		\setlength{\tabcolsep}{1.5mm}{
			\begin{tabular}{c c c c} 
				\toprule
				\textbf{Method}&\textbf{STOI} &\textbf{ESTOI}&\textbf{PESQ}\\ 
				\midrule
				End-to-End GAN \cite{mira2021end} & 0.568 &0.289 & 1.370\\
				Vocoder-based \cite{michelsanti2020vocoder} & 0.537 &	0.227&	1.230 \\
				VCA-GAN \cite{kim2021lip} & 0.569& 0.336 & 1.372 \\
				SVTS \cite{de2022svts} & 0.588 & 0.318 & 1.400 \\
				Proposed\_Speech* & \textbf{0.589} & \textbf{0.373} & \textbf{1.437} \\
				
				\bottomrule
			\end{tabular}
		}
	\end{center}
  \vspace{-2em}
\end{table}

\vspace{-1mm}
\subsection{Analysis}
\vspace{-1mm}
To verify the disentangling effect of the VAE structure, we visualized the outputs of the video content encoders in \textbf{Proposed-VAE} and  \textbf{Proposed} for unseen speakers.
For each utterance, the sequence of content representations were converted to a vector by average pooling.  The t-SNE \cite{van2008visualizing} visualization of the vectors is shown in the first column of Fig.~\ref{visual}. 
It is clear that the outputs of the video content encoder in \textbf{Proposed-VAE} contained rich speaker information and those in \textbf{Proposed} were speaker-independent.
Furthermore,  we extracted speaker embeddings from the speech synthesized by the two systems for unseen speakers using Resemblyzer\footnote{https://github.com/resemble-ai/Resemblyzer}, and their  t-SNE visualization is shown in the second column of Fig.~\ref{visual}. For  \textbf{Proposed-VAE}, the speaker embeddings of both genders distributed with large overlap, while a clear boundary between two genders can be found for the \textbf{Proposed} system. 

There have been some studies on the GRID dataset that extracted speaker embeddings from natural reference speech for unseen speakers \cite{mira2021end,michelsanti2020vocoder,kim2021lip,de2022svts}.
Although this task is not the focus of this paper, we also compared the performance of our proposed method when using the speech identity encoder for voice control with these studies.
For fair comparison, we built another \textbf{Proposed\_Speech*} model following the unseen setting \cite{de2022svts} of previous studies that used 15, 8 and 10 subjects for training, validation and test, respectively.
The results are shown in Table \ref{table2}, which shows that our method outperformed all other methods on all the three objective metrics and further proves the effectiveness of our VAE-based model.

\vspace{-1mm}
\section{Conclusion}
\vspace{-1mm}
This paper has proposed a method of zero-shot Lip2Speech synthesis method with face based voice control. 
Associated voice-face representation learning enables the face identity encoder to extract speaker voice characteristics from the face images of target speakers. Then, given an FSE vector, our method synthesizes the speech matching with the speaker's face 
for both seen or unseen speakers using a VAE-based architecture. Our future work will be improving the performance of Lip2Speech under large-vocabulary conditions and investigating the cross-modal pretraining methods for Lip2Speech synthesis.

\newpage
\bibliographystyle{IEEEbib}
\bibliography{strings,refs}

\end{document}